\documentclass[12pt]{article}
\usepackage{fullpage}

\begin{document}

\begin{center}
\Large{\bf Homi Jehangir Bhabha : Architect of Modern Science and
Technology in India} \\
\bigskip\bigskip
\large{Virendra Singh$^\star$} \\
\bigskip
Tata Institute of Fundamental Research \\
Homi Bhabha Road, Mumbai 400 005, India 
\end{center}
\bigskip

\begin{center}
\underbar{Abstract}
\end{center}
\bigskip\bigskip

After describing Bhabha's early life at Bombay, now Mumbai, we discuss
his research career at Cambridge, where he made many distinguished
contributions to positron physics, cosmic rays and the meson
theory.  These include theory of positron-electron scattering (Bhabha
scattering), Bhabha-Heitler theory of cosmic ray showers and
prediction of heavier electrons (ie. muons).  Later in his life, after
1945, Bhabha worked in India at Bangalore and Mumbai.  In India Bhabha
laid foundations of modern nuclear science and technology.  He emerged
as a successful institution builder founding Tata Institute of
Fundamental Research at Mumbai and the Laboratories of Atomic Energy
Establishment at Trombay, now renamed as Bhabha Atomic Research
Center.

\vfill

\noindent $^\star$ email: ~vsingh1937@gmail.com,

\hspace*{1cm} vsingh@theory.tifr.res.in

\newpage

\noindent {\bf Introduction}
\medskip

Homi Jehangir Bhabha (1909-66) started his career as a theoretical
physicist at Cambridge in the nineteen thirties and distinguished
himself by his researches in the emerging areas of high energy physics
and cosmic rays.  Later he excelled as a builder of institutions in
India devoted to modern science and technology.  He founded the Tata
Institute of Fundamental Research (TIFR) in 1945 at Bombay, a premier
institution devoted to excellence and the pursuit of research in the
frontier areas of science.  He was the motive force behind the
creation of the Atomic Energy Commission (AEC) by the Government of
India in 1948 and became its first Chairman.  When the Department of
Atomic Energy (DAE) of the Government of India was set up in 1954, he
was appointed as its Secretary.  The AEC and DAE were responsible for
establishing a chain of research laboratories, including the Atomic
Energy Establishment at Trombay (later renamed Bhabha Atomic Research
Center) and for commissioning India's nuclear reactors for research
and for the generation of power.  These activities also led to the
growth of electronics technology in the country and, somewhat later,
to those related to space technology.  Bhabha, more than any other
person, was responsible for introducing and nurturing the growth of
modern nuclear science and technology in India.  He was a multifaceted
personality equally at home in the world of arts.
\bigskip

\noindent {\bf 1. EARLY LIFE}
\medskip

Homi Bhabha was born in 1909 at Bombay in an established Parsi family
with a modern and nationalistic outlook.  The Parsis are a small
community, mostly settled around Bombay and in the nearby State of
Gujarat, and have acted as a pace setter in the process of social
change initiated by the impact of western civilization on India
brought about by the colonial rule by Britain.$^1$  His father,
Jehangir H. Bhabha, had been educated at Oxford and studied law at
Lincoln's Inn.$^2$

At New College at Oxford J.H. Bhabha was one of the founder members of
an India group `Nav Ratan' (nine jewels) which eventually transformed
into the `Indian Majilis'.  J.H. Bhabha was initially associated with
the judicial service of the State of Mysore.  He shifted to Bombay in
1908 after his engagement to Miss Mehram Pandey who came from the
business family of the Petits and was a granddaughter of Sir Dinshaw
Maneckji Petit, first Indian baronet and a philanthropist.  He joined
the Bar at Bombay and was for a while Presidency Magistrate.  Later,
in 1915, he joined the firm of Tata Sons and was directly involved in
their scheme of hydroelectric power generation.  He was also a trustee
of the J.N. Tata Trust and Lady Meherbai Tata Education Trust and
other philanthropic schemes.  As a young man, Homi Bhabha was exposed
to discussions on large industrial projects involving hydroelectric
power, steel and chemicals.  Homi Bhabha's grandfather, Dr. Hormusji
J. Bhabha, CIE, had been Inspector-General of Eduction with the
Education Service of the State of Mysore.  Homi Bhabha had available
to him the large library of books left behind by his grandfather.  To
this library had been added a large collection of books on art and
recorded western music on gramophone discs by his father.

Homi Bhabha's aunt Meherbai, the only sister of his father, was
married to Sir Dorabji Tata.  He used to go to his aunt's place
regularly for lunch during his school days at Cathedral and John
Connon High School at Bombay as the school was situated quite close to
the home of the Tatas.  The Lady Dorab Trust was later to play a
significant role in Homi Bhabha's life.$^3$

Bhabha left for Cambridge to study engineering in 1927 and joined
Gonville and Caius College.  Sir Dorab Tata had also spent two years
at the college and had donated 25,000 pounds to it in 1920.
\bigskip

\noindent {\bf 2. BHABHA AT CAMBRIDGE}
\medskip

The Cambridge of that period was an exciting place to be in for a
physicist.  It was there, in 1927, that Dirac had formulated the
quantum theory of emission and absorption of electromagnetic
radiation, i.e. quantum electrodynamics.  He followed it next year
with his celebrated Dirac equation for the electron, which provided a
natural explanation for the spin and the magnetic moment of the
electron and for the Sommerfeld formula of the energy levels of the
hydrogen atom.  The new feature of the Dirac equation corresponding to
`negative energy electrons' was brilliantly vindicated by the
discovery of positrons in cosmic rays by C.D. Anderson in 1932.
Bhabha was caught by this excitement and wrote to his father on 8
August, 1928 for permission to shift from engineering to physics.  He
wrote, `I seriously say to you that business or a job as an engineer
is not the thing for me.  It is totally foreign to my nature and
radically opposed to my temperament and opinions.  Physics is my line.
I know I shall do great things here'.$^4$  His father was willing to
support a further stay of two years at Cambridge to let him pursue
theoretical physics and get a mathematics tripos provided Bhabha first
devoted himself to his engineering tripos and got a first class.  In
June 1930 he passed his mechanical tripos in the first class and was
now free to devote himself to theoretical physics.  It is amusing
that, contrary to what Bhabha wrote to his father, he was to be a
great success in managing large enterprises in science and technology
and his engineering background was also to stand him in good stead.
The bedrock of acceptability of his leadership was, however, provided
by his achievements in physics.  In the milieu of Cambridge he
flowered into a fine physicist.

His earliest research interest was in quantum electrodynamics,
especially with processes involving positrons.  In 1935 he worked on
`The scattering of positrons by electrons with exchange on Dirac's
theory of positrons'.  This process in now known as Bhabha Scattering
and was a confirmation of the view that negative energy electrons can
indeed by interpreted as positrons.  Bhabha and Heitler's work in 1937
provided a natural explanation of the observed phenomenon of cosmic
ray showers.  A high energy electron traversing matter loses energy by
producing hard quanta of radiation and these quanta again materialize
into high energy electron-positron pairs.  This conversion of energy
into radiation and reconversion into secondary charged particle pairs
takes place repeatedly, resulting in an `effective' shower loss of
energy by the high energy charged particles.  Before their work it was
suspected that this `effective' shower loss of energy of the primary
particle signifies a breakdown of quantum electrodynamics.  Their work
completely accounted for the properties of the `soft' component of
cosmic rays and thus showed that no failure of the standard theory was
involved.  The studies on the absorption of cosmic rays in lead had
suggested that, apart from the `soft' component, cosmic rays also had
another component, the `penetrating component'.  A beautiful
phenomenological analysis of the data on this component of cosmic rays
led Bhabha to postulate the existence of new particles with positive
and negative electronic charge and of mass of the order of a hundred
times the mass of an electron.  Bhabha thus predicted the existence of
muons which were almost contemporaneously discovered experimentally by
Anderson.  Initially there was of course some confusion between nuons
and the particles, now called pions, postulated by Yukawa to account
for nuclear forces.  Bhabha was the first to suggest that these
particles with mass intermediate between electrons and protons,
`mesons', must spontaneously decay and that Einstein time dilation
formula can be experimentally tested by measuring the variation of the
life times of these `mesons' with their velocity in the laboratory
frame.  He, together with a select band of British scientists,
N. Kemmer, H. Frohlich and W. Heitler, was among the earliest major
workers on meson theory outside Yukawa's group in Japan.  In fact the
word `meson' was coined in a discussion between him, Kemmer and
Pryce.$^5$ 
\bigskip

\noindent {\bf 3. BHABHA IN INDIA}
\medskip

\noindent {\bf 3.1. AT BANGALORE}
\medskip

In 1939 when Bhabha was in India on a holiday visit, the second world
war broke out in Europe.  As a result of uncertain war conditions
prevailing there he could not go back to Cambridge and found himself
stranded in India.  He joined the Indian Institute of Science, in
1940, as Reader in the Physics Department in charge of the Cosmic Ray
Research Unit set up by the Sir Dorabji Tata Trust for him.
Incidentally the Trust had also provided a cyclotron to M.N. Saha
around the same time.  While at Bangalore he was elected Fellow of the
Royal Society in 1941 and was also promoted to Professor the following
year. 

During the Bangalore period (1940-45) Bhabha initiated work on
classical relativistic spinning point particle theory (Bhabha-Corben
equations), and continued working on meson theory, leading to the
first suggestion about the existence of nucleon isobars, and on
cascade theory.  He also initiated experimental work in the field of
cosmic rays.  Apart from scientific research work `he found his
mission in life'$^6$ during this period.  He had been away from India
and exposed to the scientific and cultural life of Europe for thirteen
years, leaving India at the young age of seventeen to return at the
age of twenty-nine.  He was now exposed to India as a mature young
man, and as a result of introspection and variety of experiences
during this period, he began to identify himself more and more with
India and its cultural heritage.  He saw that there was much work to
be done here if science and technology were to be harnessed for the
development of the country.  He saw his mission  as bringing modern
science and technology to India.
\bigskip

\noindent {\bf 3.2. TATA INSTITUTE OF FUNDAMENTAL RESEARCH}
\medskip

On 12 March 1944 Bhabha wrote a formal letter proposing the setting up
of an Institute of Fundamental Research in India to Sir Sorab
Saklatvala, Chairman of the Sir Dorabji Tata Trust.  In this he noted,
`I had the idea that after the war I would accept a job in a good
university in Europe or America, because universities like Cambridge
or Princeton provide an atmosphere which no place in India provides at
the moment.  But in the last two years I have come more and more to
the view that provided proper appreciation and financial support are
forthcoming, it is one's duty to stay in one's own country and build
up schools comparable with those that other countries are fortunate in
possessing ....  The scheme I am now submitting to you is but an
embryo from which I hope to build up in the course of time a School of
Physics comparable with the best anywhere'.  Before this formal letter
there had been exploratory correspondence with J.R.D. Tata and with
his encouragement and support the Sir Dorabji Tata Trust decided to
support the scheme.  The proposed institute, named Tata Institute of
Fundamental Research, started functioning with Bhabha as its founder
director in June 1945 at Bangalore.  It was soon shifted to Bombay and
was formally inaugurated in its temporary premises at Kenilworth
(Peddar Road) on Wednesday, 19 December 1945.  He had indicated in the
same letter his initial choice of research areas for the institute,
`The subjects on which research and advanced teaching would be done
would be theoretical physics, especially on fundamental problems and
with special referece to cosmic rays and nuclear physics, and
experimental research on cosmic rays'.  Bhabha worked on his theory of
relativistic wave equations for elementary particles (Bhabha wave
equations), cascade theory and stochastic processes, and on multiple
meson production.  His last research paper in theoretical physics was
written in 1953.  Later he became more and more involved with
developing the atomic energy programme in India.
\bigskip

\noindent {\bf 3.3. ATOMIC ENERGY}
\medskip

Already at the time of writing the letter to Sir Dorabji Tata Trust on
12 March 1944, Bhabha had visualized that the proposed institute
devoted to basic research should play a role in developing high
quality applied research and nuclear power programmes in India.  As he
wrote, `There is at the moment in India no big school of research in
the fundamental problems of physics, both theoretical and
experimental.  There are, however, scattered all over India competent
workers who are not doing as good work as they would do if brought
together in one place under proper direction.  It is absolutely in the
interest of India to have a vigorous school of research in fundamental
physics, for such a school forms the spearhead of research not only in
less advanced branches of physics but also in problems of immediate
practical application in industry.  If much of the applied research
done in India today is disappointing or of very inferior quality it is
entirely due to the absence of a sufficient number of outstanding pure
research workers who would set the standard of good research and act
on the directing boards in an advisory capacity. ....Moreover, when
nuclear energy has been successfully applied for power production in
say a couple of decades from now, India will not have to look abroad
for its experts but will find them ready at hand'.  It has been
remarked often that this letter suggesting peaceful nuclear
applications was written more than a year before the world at large
became aware of atomic energy unfortunately through its military
application at Hiroshima and Nagasaki.

An Atomic Energy Committee was constituted by the Council of
Scientific and Industrial Research in India towards the end of 1945
with Bhabha as the Chairman.  Some of the other members were
S.S. Bhatnagar, M.N. Saha and D.N. Wadia.  The committee initiated
some survey work on atomic minerals.  On 26 April 1948 Bhabha
submitted to the Prime Minister, Jawaharlal Nehru, a comprehensive
`note on the organization of atomic research in India' and recommended
the formation of the Atomic Energy Commission of India.$^7$  The
Commission was established in August 1948 with Bhabha as Chairman, and
S.S. Bhatnagar and K.S. Krishnan as members.  Its charter requested
the Atomic Energy Commission.  `(1) to take such steps as may be
necessary from time to time to protect the interests of the country in
connection with Atomic Energy by exercise of the powers conferred on
the Government of India by the provisions of the Atomic Energy Act;
(2) to survey the territories of the Indian Dominion for the location
of useful minerals in connection with Atomic Energy; and (3) to
promote research in their own laboratories and to subsidize such
research in existing institutions and universities.  Special steps
will be taken to increase teaching and research facilities in nuclear
physics in the Indian Universities'.

The initial trained manpower for atomic energy work came from the Tata
Institute of Fundamental Research.  The Physics Division of AEC was
housed at TIFR in the early days.  The early activities of the
electronics and technical physics division of the Atomic Energy
Establishment at Trombay (later renamed Bhabha Atomic Research Centre)
were also started at TIFR.  This electronics production unit at TIFR
eventually developed into the Electronics Corporation of India Ltd.
The control systems for Apsara, which was Asia's first nuclear
reactor, were also built at TIFR.  As Bhabha mentioned in his speech
at the inaugural function of the new building for TIFR at its present
location on 15 January 1962, `It is not an exaggeration to say this
institute was the cradle of our atomic energy programme, and if the
Atomic Energy Establishment at Trombay has been able to develop so
fast, it is due to the assisted take-off which was given to it by the
institute in the early stages of its development.  It is equally true
to say that the Institute could not have developed to its present size
and importance but for the support it has received from the Government
of India'.

By the middle of 1949 the Rare Minerals Survey Unit, later renamed
Atomic Minerals Division, had been set up.  The Indian Rare Earths
Ltd. was started in 1950 to mine and process Kerala and Orissa beach
sands.  The Government of India created the Department of Atomic
Energy in 1954 with Bhabha as its Secretary.  It is to be remarked
that this decision to create a department of atomic energy owes much
to the vision and faith of Bhabha and Nehru as, at that time, use of
atomic energy to generate electricity was not a proven technology even
in the developed Western countries.  The swimming pool type research
reactor, Apsara, achieved criticality on 4 August 1956.  The research
laboratories of the Atomic Energy Establishment at Trombay were
formally inaugurated by Prime Minister Nehru on 20 January 1957 and
produced nuclear grade uranium metal in 1959.  The reactor CIRUS and
the research reactor ZERLINA became critical in 1960 and 1961
respectively.  The plutonium plant was commissioned in 1964.  The
Power Projects Engineering Division was created to devote itself
entirely to the Indian nuclear power programme.

Following the initiative of President Eisenhower of USA, it was
decided by the United Nations in December 1954 to hold an
international conference on developing the peaceful uses of atomic
energy in Geneva in 1955.  Bhabha presided over this first United
Nations conference with grace and distinction.  His presidential
address on the contributions which nuclear energy could make to solve
the world energy supply problems was masterly.  He was, however,
unduly optimistic about the prospects of the exploitation of energy
from nuclear fusion.
\bigskip

\noindent {\bf 3.4. ELECTRONICS AND SPACE PROGRAMME}
\medskip

Apart from pioneering atomic energy in the country, Bhabha also played
a crucial role in introducing space research and programme in India.
Largely through his initiative, the Indian National Committee for
Space Research was set up in 1962 under the chairmanship of Vikram
Sarabhai.  The first rocket launch was carried out from Thumba
Equatorial Rocket Launching Station near Trivandrum on 21 November
1963.  Bhabha also announced on that day the formation of a space
science and technology centre.  The Thumba facility was made available
to the international community through the United Nations in December
1965.  We have noted how the early work in electronics and
instrumentation laid the foundations for later work at atomic energy
laboratories and to the establishment of the Electronics Corporation
of India.  Bhabha gave a high priority to the development of
electronics in India.  He therefore accepted the Chairmanship of the
Electronic Committee constituted by the Government of India.  The
Bhabha Committee report made recommendations concerning the future
plans and development for the electronics industry in India.

Bhabha tragically died in an air crash on Mont Blanc on 24 January
1966.  He was on his way to attend a meeting of the International
Atomic Energy Agency at Vienna.  Indian science and technology lost a
very distinguished man of vision and drive at a crucial juncture of
its post-independence development.
\bigskip

\noindent {\bf 4. BHABHA AS A RENAISSANCE MAN}
\medskip

Bhabha was likened to `a man of renaissance'.$^8$  He was a man of
science and technology.  He was a great administrator of science.  He
was equally at home in the world of arts.  Bhabha grew up listening to
his father's and his aunt's excellent collection of recorded western
classical music.  Later he developed a taste for classical Indian
music and dance also.  The choice of Vienna as the headquarters of the
International Atomic Energy Agency was to a large extent decided by
Bhabha's desire to combine attendance at its meetings with an
opportunity to attend the fine musical concerts there.  He enjoyed
sketching and painting and left behind a number of these.$^9$  As
M.F. Hussain noted, `Though a scientist by profession, he was an
artist by nature .... .  To a great extent Bombay is what it is today
because of Bhabha.  In the early fifties he conceived the idea of
starting a collection of painting at TIFR .... and Bombay witnessed
the birth of modern art'.$^{10}$  He was also uniquely qualified to
edit a special issue of the Indian art magazine Marg on the occasion
of the five hundredth anniversary of Leonardo da Vinci,$^{11}$ the
original renaissance man.  He designed the buildings of the Tata
Institute of Fundamental Research with great care and attention to
every minute detail.  The glass windows had frames of a special new
aluminium alloy to withstand sea breeze corrosion, and this alloy was
manufactured for the first time in India at Bhabha's initiative.  The
buildings are not only functional but also beautiful.  As Helmut
Bartsch, the architect, noted, `In this development the architect
worked with a client rather than for a client.  The client displayed
unending interest and encouragement and constantly added intelligent
suggestions and advice.  The result, it is hoped, is a building which
will not only fulfil its function but should afford a great deal of
enjoyment'.$^{12}$  The siting of the Atomic Energy Establishment is
such that the island of Elephanta, which contains beautiful seventh
century cave temples, is directly visible.  The same juxtaposition
occurs between Kalpakkam reactor centre and the famed Mahabalipuram
temples.  He also had a keen sense of landscape and the Tata Institute
of Fundamental Research and the Bhabha Atomic Research Centre are
surrounded by beautiful trees and gardens.$^{13}$
\bigskip

\noindent {\bf 5. BHABHA'S WRITINGS}
\medskip

Bhabha brought to India nuclear science and technology and left behind
him a legacy of large and successful scientific institutions,
technological laboratories and power reactors.$^{14}$  He contributed
a great deal to inculcate the scientific temper in India through his
activities.  He, however, did all this by his example and by creating
the needed infrastructure.  He did not do so through writing about it.
In the present selection we have tried to present some of his writings
which are close to humanist concerns.

Bhabha's research papers deal with high energy physics and cosmic
rays.  These are now available in a volume, {\it Homi Jehangir Bhabha:
Collected Scientific Papers}.$^{15}$  No collection of his other
writings and speeches has been published so far.  His major concern in
these is with atomic energy in the context of the Indian situation.
Most of these writings on atomic energy are also, by and large, of a
technical nature, involving detailed consideration of feasibility and
economics of atomic energy and full of graphs, tables and figures.
International aspects of atomic energy are also examined.  In their
style Bhabha's background as a theoretical physicist and as one given
to precision, shows through.  However, any selection from Bhabha's
writings cannot be representative without some of his papers on energy
and atomic energy. We have chosen two of the non-technical ones.  One
of these is his presidential address at the International Conference
on Peaceful Uses of Atomic Energy at Geneva in 1955.$^{16}$  As Lord
Penney says in his memoir on Bhabha, this `was a brilliant essay about
energy and population, and described in simple direct style the
changes in living conditions.  The address is as vivid and true today
as when it was written'.$^{17}$  This article puts atomic energy in
the perspective of human use of energy through its history.  In the
more specific context of developing countries he presents the case for
atomic energy in his article `The Role of Atomic Energy in Asia'
written for the inaugural issue of {\it Asian Atomic Newsletter}
brought out by the Philippines Atomic Energy Commission in
1964.$^{18}$  We have chosen this presentation as it was one of his
few general writings.

Bhabha's broad views on science and technology and its place in the
world is best brought out in a paper `The Role of Science and
Technology in Producing an Integrated World' which he prepared for the
Centennial Conference on Science and Engineering Education at
Massachusetts Institute of Technology in 1961.  A lecture delivered at
a meeting of the International Council of Scientific Unions `Science
and the Problems of Development'$^{19}$ on 7 January 1966 provides an
authoritative version of the philosophy which Bhabha followed in his
activities as a builder of institutions towards the development of
science and technology in India.$^{20}$  It is as thought-provoking
and relevant today as when it was delivered.

The author is thankful to Professor B.V. Sreekantan and Dr. A.K. Raina
for their useful comments on the manuscript.

\newpage

\begin{center}
{\bf REFERENCES AND FOOTNOTES}
\end{center}

\begin{enumerate}
\item[{1.}] E. Kulke, {\it The Parsis in India: A Minority as Agent of
Social Change}, Vikas Pub., New Delhi 1978.  
\item[{2.}] A biographical sketch, a photograph and obituary notices
of Bhabha's father occur in `The Late Mr. J.H. Bhabha', {\it
Electrotechnics}, No. 15-16 (August 1946).
\item[{3.}] For an account of the first fifty years of Sir Dorabhji
Tata Trust and its founder see R.M. Lala, {\it The Heartbeat of a
Trust}, Tata McGraw-Hill Pub. Co., New Delhi 1984.  Homi Bhabha and
Tata Institute of Fundamental Research are dealt in Chapters 8-10
(pages 25-124).
\item[{4.}] Ibid. 3, p.85.
\item[{5.}] For more detailed account of his scientific work see
B.V. Sreekantan, `H.J. Bhabha: His Contribution to Cosmic Ray
Physics'; V. Singh, `H.J. Bhabha: His Contribution to Theoretical
Physics', in B.V. Sreekantan, Virendra Singh and B.M. Udgaonkar
(eds.), {\it H.J. Bhabha: Collected Scientific Papers},
pp. xi-xxii,xxiii-xlvii. 
\item[{6.}] M.G.K. Menon in {\it Homi Jehangir Bhabha 1909-1966}, The
Royal Institution of Great Britain, London 1967, p.17.
\item[{7.}] This note and some other documents relating to Bhabha's
work on atomic energy and his selected correspondence with Jawaharlal
Nehru appear in a special issue, {\it Nuclear India}, vol. 26,
no. 10(1989), compiled by M.V. Ramaniah. 
\item[{8.}] C.F. Powell, `Homi Bhabha-A man of renaissance', {\it
Science Reporter}, vol. 3, no. 10 (Oct. 1966).  This issue of {\it
Science Reporter}, devoted to Bhabha, also contains tributes to him
from Indira Gandhi, J.B. Wiesner, B. Goldschmidt, S. Eklund,
B. Peters, W. Heitler, W.B. Levis, J.R.D. Tata, R. von Leyden and
others. 
\item[{9.}] Some of Bhabha's drawings and sketches appear in (i) {\it
Homi Bhabha as Artist: A Selection of his Paintings, Drawings and
Sketches}, edited by Jamshed Bhabha (Marg Pub., Nov. 1968), (ii) {\it
Homi Bhabha: A Portfolio of Drawings} (Marg Publication).
\item[{10.}] M.F. Husain, Interview in MEGA, October 1987.
\item[{11.}] H.J. Bhabha, `Leonardo Da Vinci, Five Hundred Years',
Marg. 5, \#4, 1 (1952). 
\item[{12.}] Notes by Helmut Bartsch in `The Tata Institute of
Fundamental Research, Inauguration of New Buildings', Bombay, 15
January 1962, p.51.
\item[{13.}] S.D. Vaidya, `For Whom the Trees Grieve', {\it Science
Today}, vol. 18, no. 10 (Oct. 1984).
\item[{14.}] R.S. Anderson, `Building Scientific Institutions in
India: Saha and Bhabha' Occasional Paper series no. 11, Centre for
Developing Area Studies, McGill University, Montreal 1975.
\item[{15.}] {\it Homi Jehangir Bhabha, Collected Scientific Papers},
Edited by B.V. Sreekantan, Virendra Singh and B.M. Udgaonkar, Tata
Institute of Fundamental Research, Bombay 1985.
\item[{16.}] H.J. Bhabha, Presidential address, {\it
Proc. Int. Conference on Peaceful Uses of Atomic Energy} (Geneva),
United Nations, NY, 1, 103 (1956).    
\item[{17.}] Lord Penney, `Homi Jehangir Bhabha (1909-1966)', {\it
Biographical Memoirs of Fellows of the Royal Society}, 13, 35-55
(1967), p.49.
\item[{18.}] H.J. Bhabha, `Role of Atomic Energy in Asia', {\it Asian
Atomic Newsletter}, 1, 1 (1964).
\item[{19.}] This lecture was also reprinted in (i) `H.J. Bhabha,
Indian Science -- Two Methods of Development', {\it Science and
Culture} (July 1966), pp. 333-342, (ii) {\it Physics News} 19, 100
(1988). 
\item[{20.}] A selection of quotations from Bhabha about his
philosophy of growth of science is available in B.M. Udgaonkar, `Homi
Bhabha on Growing Science', in {\it H.J. Bhabha, Collected Scientific
Papers}, xlix-lxxix.
\end{enumerate}

\end{document}